\PassOptionsToPackage{table,xcdraw}{xcolor} 
\documentclass[sigconf]{acmart}
\usepackage{multirow} 
\usepackage{subfigure}
\usepackage{booktabs}
\usepackage{makecell}
\usepackage{caption}
\usepackage{amsmath}
\usepackage[table,xcdraw]{xcolor}
\usepackage{soul}
\usepackage{array}
\usepackage{geometry}
\usepackage{colortbl}
\usepackage{url}
\usepackage{soul}

\usepackage{arydshln}

\definecolor{lightblue}{RGB}{221, 231, 245} 
\definecolor{lightyellow}{RGB}{209, 239, 241}
\definecolor{lightgreen}{RGB}{255, 240, 230}
\definecolor{lightred}{RGB}{255,102,102}

\AtBeginDocument{%
  }

\copyrightyear{2025}
\acmYear{2025}
\setcopyright{acmlicensed}\acmConference[SIGIR '25]{Proceedings of the 48th International ACM SIGIR Conference on Research and Development in Information Retrieval}{July 13--18, 2025}{Padua, Italy}
\acmBooktitle{Proceedings of the 48th International ACM SIGIR Conference on Research and Development in Information Retrieval (SIGIR '25), July 13--18, 2025, Padua, Italy}
\acmDOI{10.1145/xxxxxxx.xxxxxxx}
\acmISBN{979-x-xxxx-xxxx-x/xxxx/xx}

\begin{document}

\title{Enhancing the Patent Matching Capability of Large Language Models via the Memory Graph}

\author{Qiushi Xiong}
\authornote{ \ \ indicates equal contribution.}
\affiliation{%
  \institution{Northeastern University}
  \city{Shenyang}
  \country{China}
}
\email{xiongqiushi@stumail.neu.edu.cn}

\author{Zhipeng Xu}
\authornotemark[1]
\affiliation{%
  \institution{Northeastern University}
  \city{Shenyang}
  \country{China}
}
\email{xuzhipeng@stumail.neu.edu.cn}

\author{Zhenghao Liu}
\authornote{ \ \ indicates corresponding author.}
\affiliation{%
  \institution{Northeastern University}
  \city{Shenyang}
  \country{China}}
\email{liuzhenghao@mail.neu.edu.cn}

\author{Mengjia Wang}
\affiliation{%
  \institution{Alibaba Group}
  \city{Hangzhou}
  \country{China}
}
\email{mengjia.wmj@alibaba-inc.com}

\author{Zulong Chen}
\affiliation{%
  \institution{Alibaba Group}
  \city{Hangzhou}
  \country{China}
}
\email{zulong.czl@alibaba-inc.com}

\author{Yue Sun}
\affiliation{%
  \institution{Alibaba Group}
  \city{Hangzhou}
  \country{China}
}
\email{sy208806@alibaba-inc.com}

\author{Yu Gu}
\affiliation{%
  \institution{Northeastern University}
  \city{Shenyang}
  \country{China}}
\email{guyu@mail.neu.edu.cn}

\author{Xiaohua Li}
\affiliation{%
  \institution{Northeastern University}
  \city{Shenyang}
  \country{China}}
\email{lixiaohua@mail.neu.edu.cn}

\author{Ge Yu}
\affiliation{%
  \institution{Northeastern University}
  \city{Shenyang}
  \country{China}}
\email{yuge@mail.neu.edu.cn}

\renewcommand{\shortauthors}{Qiushi Xiong et al.}


\begin{abstract}
Intellectual Property (IP) management involves strategically protecting and utilizing intellectual assets to enhance organizational innovation, competitiveness, and value creation. Patent matching is a crucial task in intellectual property management, which facilitates the organization and utilization of patents. Existing models often rely on the emergent capabilities of Large Language Models (LLMs) and leverage them to identify related patents directly. However, these methods usually depend on matching keywords and overlook the hierarchical classification and categorical relationships of patents. In this paper, we propose \method{}, a method that augments the patent matching capabilities of LLMs by incorporating a memory graph derived from their parametric memory. Specifically, \method{} prompts LLMs to traverse their memory to identify relevant entities within patents, followed by attributing these entities to corresponding ontologies. After traversing the memory graph, we utilize extracted entities and ontologies to improve the capability of LLM in comprehending the semantics of patents. Experimental results on the PatentMatch dataset demonstrate the effectiveness of \method{}, achieving a 17.68\% performance improvement over baseline LLMs. The further analysis highlights the generalization ability of \method{} across various LLMs, both in-domain and out-of-domain, and its capacity to enhance the internal reasoning processes of LLMs during patent matching. All data and codes are available at \url{https://github.com/NEUIR/MemGraph}.
\end{abstract}





\def\method{MemGraph}

\maketitle
\begin{figure}[t]        
    \centering
    \includegraphics[width=0.47\textwidth]{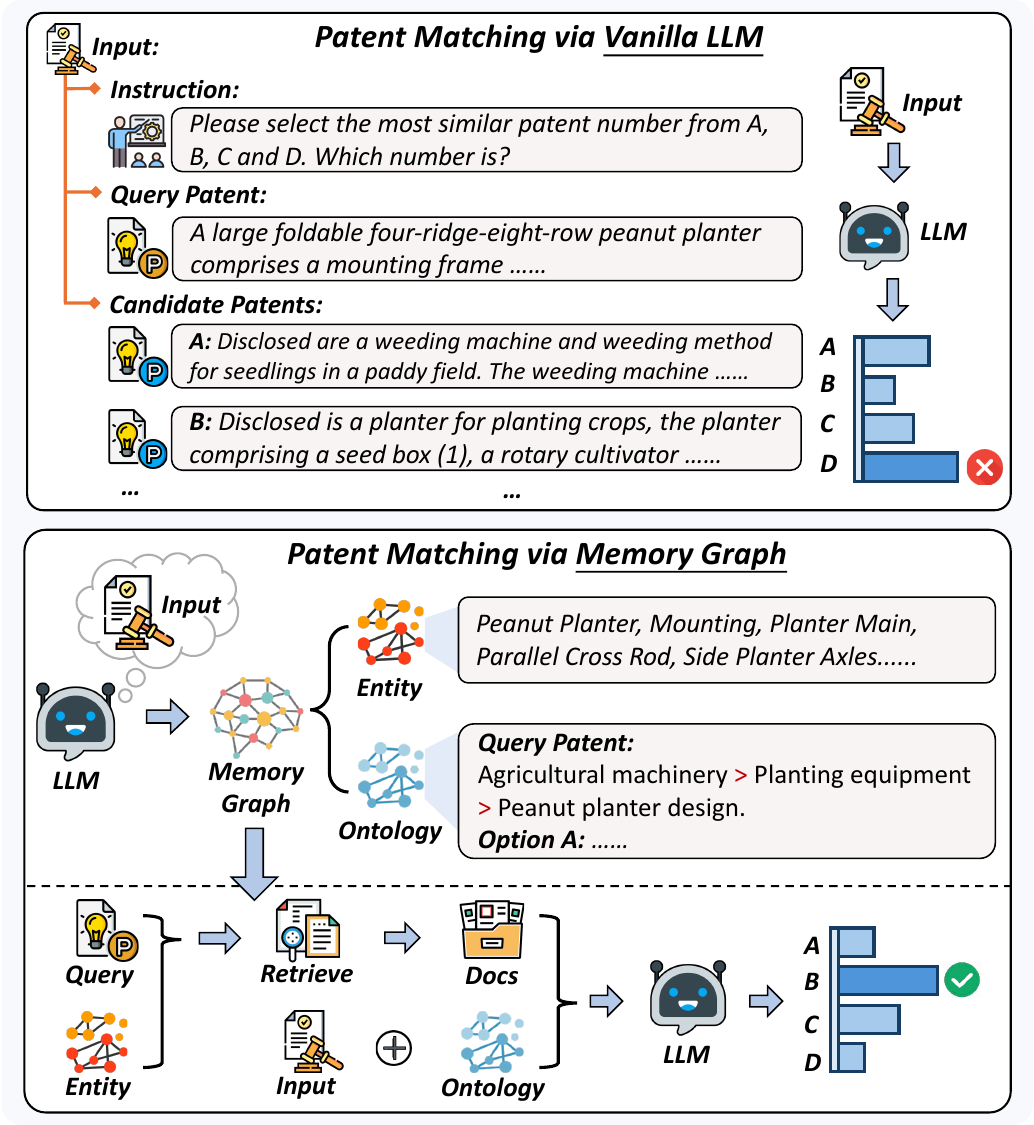}
    \caption{\label{fig:intro}Illustration of Our \method{} Method. The framework integrates memory graph into LLM-based patent matching, enabling more comprehensive semantic understanding and accurate patent similarity assessment.}
\end{figure}

\section{Introduction}
Patent matching is a critical aspect of intellectual property (IP) management, enabling the identification of similar patent documents with overlapping innovations and aiding in the prevention of potential infringements~\cite{abraham2001innovation,arts2018text,krestel2021survey}. However, the intricate structures of patent documents and the rapid evolution of specialized terminology driven by technological advancements, also pose significant challenges~\cite{peng2024connecting}. These complexities make it difficult to discern subtle differences among patents, thereby making it difficult to achieve more accurate matching results among patents~\cite{tseng2007text,xue2009transforming,lupu2017current,risch2020patentmatch}.

Large Language Models (LLMs), \textit{e.g.} ChatGPT~\cite{achiam2023gpt} and LLaMA~\cite{touvron2023llama}, have shown the potential to deal with technical knowledge extraction, patent analysis, and patent matching, facilitating the processing of patent literature~\cite{wang2024patentformer, pelaez2024large, sakhinana2024towards}. Despite their advancements,
recent findings reveal that LLMs usually face the vocabulary mismatch problem~\cite{furnas1987vocabulary, croft2010search} in domain-specific tasks~\cite{jiang2024can,siddharth2024retrieval, halgin2024intelligent}. This challenge manifests as difficulties in comprehending specialized terminologies and the technical details of inventions~\cite{xu2024activerag, ni2024mozip, jiang2024artificial}.
To alleviate the domain gap of LLMs, existing research primarily focuses on training LLMs to learn domain-specific knowledge. Specifically, these approaches pretrain LLMs on domain-specific corpora~\cite{ni2024mozip}, construct an instruct-tuning dataset for domain-specific task~\cite{ni2024mozip, bai2024patentgpt} or align LLMs’ matching behavior with that of human examiners~\cite{ren2024patentgpt}. However, such methods often require substantial amounts of domain specific data and risk catastrophic forgetting during domain-specific training~\cite{ramasesh2021effect}.
Instead of domain-specific training, recent work in patent analysis leverages Retrieval-Augmented Generation (RAG) methods~\cite{lewis2020retrieval, cai2022recent, gao2023retrieval} to update the parametric memory of LLMs by incorporating external knowledge. These approaches primarily focus on providing more fine-grained external knowledge and reducing noise in retrieved patents during the patent analysis process. In detail, they parse patent documents into graphs as a knowledge base, retrieve relevant triplets, and integrate them into the input context of LLMs~\cite{sakhinana2024towards,siddharth2024retrieval,chu2024patent}, aiming to use this graph-based knowledge to enhance LLMs' understanding of patent documents.
However, these methods often overlook the potential of LLMs to automatically parse patent literature and establish more effective relationships among patents.

This paper proposes \method{}, a novel framework to enhance the patent matching capabilities of LLMs through the memory graph derived from their parametric memory.
As illustrated in Figure~\ref{fig:intro}, our method activates the memory graph of LLMs, extends the retrieval-augmented approach, and improves both the retrieval and matching processes.
Specifically, \method{} first leverages LLMs to traverse the memory graph based on query patents, identify related entities and expand the query patents to conduct more accurate retrieval results. These entities are the keywords of the patent, such as ``Peanut Planter'' and ``Mounting''. Subsequently, \method{} traverses the memory graph to extract related ontologies for both query and candidate patents, such as ``Agricultural machinery'' and ``Planting equipment''. These ontologies encapsulate broader concepts, categories, or relationships, which are then utilized to establish the relationship between different patents. This enables LLMs to better understand the hierarchical and relational structures of patents, enhancing the patent matching process.
 
Our experimental results on the PatentMatch dataset demonstrate the effectiveness of \method{}, achieving an average accuracy improvement of 17.68\% over vanilla LLMs and 10.85\% over vanilla RAG models. Notably, \method{} consistently outperforms baseline methods when using different LLMs as backbone models, highlighting its strong generalization ability.
Further analysis reveals that \method{} enhances patent matching performance by leveraging related entities and ontologies as hints to assist both the retriever and generator within the RAG framework. 
Specifically, query-related entities enable the retriever to provide more accurate retrieved patents, while the ontologies related to both the query and candidate patents help the generator better utilize internal and external knowledge. 
By incorporating these extracted ontologies, \method{} enables LLMs to reduce prediction uncertainty and improve reasoning quality during the matching process, resulting in more convincing matching.
\section{Related Work}

Patent matching has long been a challenging task due to the complexity and diversity of patent documents~\cite{masur2010costly, king2003patent, krestel2021survey,abbas2014literature, graham2018uspto}.
Unlike patent retrieval task which focuses on finding relevant patents~\cite{ villa2022sequential, roudsari2021comparison, chikkamath2024your}, patent matching emphasizes assessing the similarity of technological innovations described in patent documents~\cite{garfield1966patent, hain2022text, ni2024mozip}. Earlier works have attempted to conduct keyword matching~\cite{indukuri2007similarity, cascini2008measuring, park2012identifying} or to learn an embedding space~\cite{ascione2024comparative, hain2022text, roudsari2021comparison} for modeling the similarity of innovations in patents. However, these methods often struggle to capture the nuanced technological innovations within similar patents, resulting in limited effectiveness~\cite{risch2020patentmatch}.

Recent research mainly focuses on using LLMs to build efficient patent matching frameworks, which thrive on leveraging their strong semantic understanding and emergent capabilities~\cite{zhao2023survey, wei2022emergent, kojima2022large}. These methods primarily focus on making LLMs learn domain-specific terminologies and concepts by training on patent data. Specifically, MoZi~\cite{ni2024mozip} continuously pretrains LLMs with a well-curated patent corpus, followed by instruction tuning on patent-related questions to enhance the capability of LLMs in analyzing the technical details of patent documents. PatentGPT~\cite{ren2024patentgpt} designs additional pretraining tasks based on patent knowledge graphs to help LLMs better capture the relationships between entities in patents. PatentGPT-1.0-Dense~\cite{bai2024patentgpt} further incorporates Reinforcement Learning with Human Feedback (RLHF) to align the matching behavior of LLMs with that of human examiners, achieving promising capabilities in the patent matching task. Nevertheless, these training methods struggle with the rapid evolution of domain-specific terminology~\cite{wen2023mindmap} and the risk of catastrophic forgetting during continuous training~\cite{ramasesh2021effect}.

To effectively update the patent knowledge of LLMs, recent work in patent analysis has primarily focused on enhancing the domain-specific capabilities of LLMs using the Retrieval-Augmented Generation (RAG) framework~\cite{lewis2020retrieval, jiang2023active, xu2024activerag, asai2023self}, rather than continuously training of the LLMs. However, several studies have highlighted that noise in retrieved documents can mislead LLMs, leading to knowledge conflicts and performance degradation~\cite{xie2023adaptive, xu2024knowledge, li2024rag}. To mitigate the impact of retrieved noise, existing research has primarily concentrated on graph-based RAG methods to refine the retrieved evidence, thereby preventing LLMs from being hindered by noisy patent texts. Instead of directly feeding LLMs with retrieved patents, these methods parse patent documents into a knowledge graph and retrieve relevant triplets to enhance LLM response generation~\cite{siddharth2024retrieval, chu2024patent, sakhinana2024towards, peng2024connecting}. While these approaches provide LLMs with more refined knowledge, they do not fully utilize the retrieved evidence to facilitate a deeper understanding of patents and fail to motivate LLMs to resolve knowledge conflicts between parametric memory and external knowledge.
\begin{figure*}[t]
    \centering
    \includegraphics[width=\textwidth]{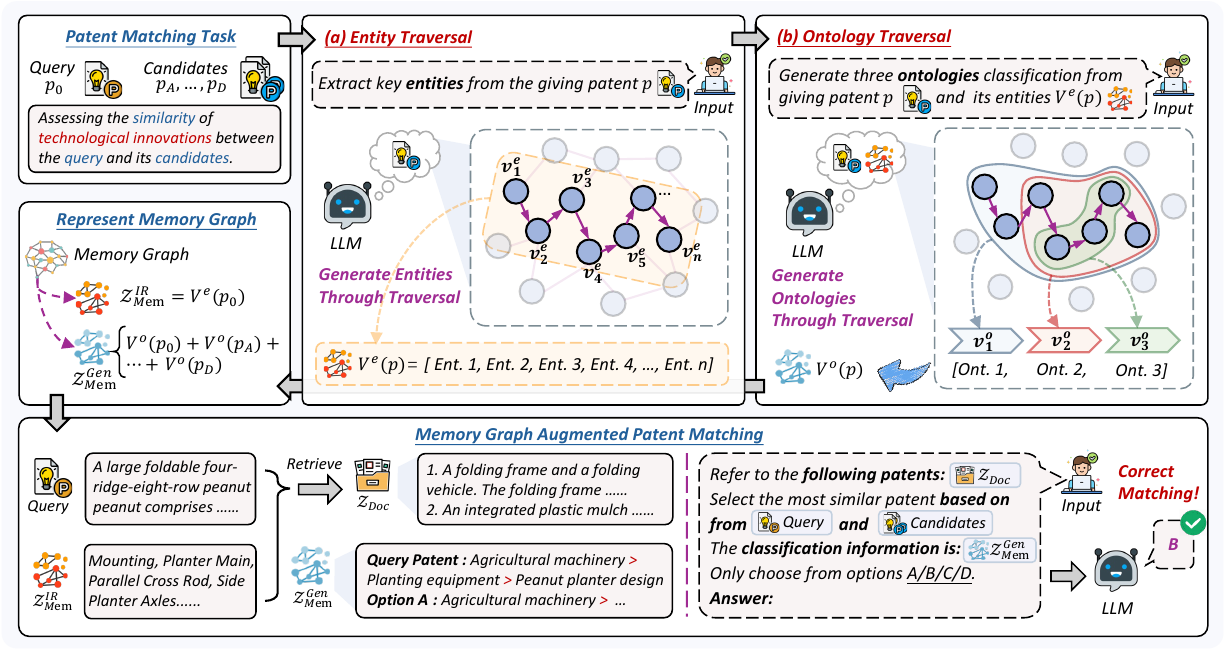} 
    \caption{\label{fig:model}The Illustration of Our \method{} Method.}
\end{figure*}
\begin{table*}[t]

\caption{Prompt Templates Used in Traversing Memory Graph for the Given Patent.}
\label{tab:prompt}
\renewcommand{\arraystretch}{1.2}
\small

\definecolor{titlecolor}{RGB}{248, 249, 253}

\setlength{\arrayrulewidth}{0.7pt}

\begin{tabular}{|p{0.98\textwidth}|}
\hline  
\cellcolor{titlecolor}\textbf{Entity Traversal} \\ \hline
You are an AI assistant specialized in patent analysis. Your task is to extract key technical entities from a given patent abstract. These entities should be specific technical concepts, components, or methods that are central to the patent's innovation. \\
\textbf{Instructions:} 1. Carefully read the provided patent abstract. 2. Identify and list the most important technical entities mentioned in the patent. 3. Focus on entities that are: - Specific to the technology described. - Central to the patent's claims or innovative aspects. - Likely to be useful for understanding the patent's technical field. 4. Provide each entity as a concise phrase or term (typically 1-5 words). 5. List up to 10 entities, prioritizing the most significant ones. 6. Do not include general or broad categories; focus on specific technical concepts. \\
\textbf{Output Format:} [Entity 1], [Entity 2], [Entity 3], ... \textbf{Patent Abstract:} \{Patent\_Abstract\} \\ \hline
\cellcolor{titlecolor}\textbf{Ontology Traversal} \\ \hline
You are a patent classification expert. In the patent examination and search process, it is often necessary to compare the technical fields of multiple related patents. I will provide an original patent abstract and its related technical entities, as well as four potentially related patent abstracts (options A, B, C, D) and their respective technical entities. The original abstract represents a patent under examination, while options A-D represent potentially related existing patents found in the database. \\
Your task is to generate a multi-level technical classification for the original abstract and each option, referencing but not limited to the approach of the International Patent Classification (IPC) system. These classifications will be used to evaluate the technical relevance between the original patent and existing patents, helping to determine the novelty and inventiveness of the patent. \\
\textbf{Input:} \{Original\_Abstract, Original\_Abstract\_Entity, Option\_A-D\_Abstract, Option\_A-D\_Abstract\_Entities\} \\
Please \textbf{output} the classification results using the following \textbf{format}: \{Type\}: [Major category] > [Subcategory] > [Specific class], where Type $\in$ \{Original Patent, Option A, Option B, Option C, Option D\} \\
\textbf{Notes:} 1. The classification should be in three levels. 2. Reference but do not limit yourself to the IPC classification system approach, using general, intuitive terms to describe technology categories. 3. Use 1-3 words to describe each level, gradually refining from major category to specific class. 4. Classifications should reflect the patent's core technical features, application areas, and innovation focus. 5. Make full use of the provided technical entities, which contain important technical information. 6. Maintain consistency: if two patents belong to similar fields, they should be given similar classifications. 7. Only output the classification results, do not add any additional explanations. \\
Based on the above information, please provide accurate and concise three-level technical classifications for the given original patent abstract and each option. Before starting the classification, please carefully read all abstracts and technical entities to ensure consistency and relevance in the classification. \\ \hline 
\end{tabular}
\end{table*}
\section{Methodology}\label{Methodology}
In this section, we first describe the preliminary of patent matching task (Sec.~\ref{model:preliminary}) and then introduce the memory graph mechanism to enhance the patent matching capability of LLMs (Sec.~\ref{model:architecture}).

\subsection{Memory Graph Augmented Patent Matching}\label{model:preliminary}
Given a query patent $p_\text{0}$, the patent matching task~\cite{ni2024mozip} focuses on identifying the patent with the most similar innovations from a set of candidates $\mathcal{P} = \{p_\text{A}, \dots, p_\text{D}\}$. Each candidate patent $p_i$ is paired with a corresponding identifier $y_i$, which is labeled as A, B, C, and D. The goal of the patent matching task is to build a framework to output the identifier $\Tilde{y}$ of the patent that most closely aligns with the innovations described in the query $p_\text{0}$:
\begin{equation}
    P(\Tilde{y}|p_\text{0}, \mathcal{P}) = \text{LLM} (p_\text{0},p_\text{A} \oplus \dots \oplus p_\text{D}),
\end{equation}
where $\oplus$ is the concatenation operation. Then, we will describe two methods to enhance the patent matching capability of LLMs by leveraging retrieved evidence and the memory graph.

\textbf{Retrieval-Augmented Patent Matching.} To enhance the accuracy of patent matching, we employ the Retrieval-Augmented Generation method, enabling LLMs to access the patent corpus $\mathcal{D}$. The probability $P(\Tilde{y}|p_\text{0}, \mathcal{P}, \mathcal{D})$ of identifying the most relevant patent can be approximately computed as:
\begin{equation}
    P(\Tilde{y}|p_\text{0}, \mathcal{P}, \mathcal{D}) \approx P(\Tilde{y}|p_\text{0}, \mathcal{P}, \mathcal{Z}_\text{Doc}),
\end{equation}
where the $\mathcal{Z}_\text{Doc}$ represents the latent variable. The probability $P(\Tilde{y}|p_\text{0}, \mathcal{P}, \mathcal{Z}_\text{Doc})$ of the most relevant patent can be calculated by LLMs:
\begin{equation}
    P(\Tilde{y}|p_\text{0}, \mathcal{P}, \mathcal{Z}_\text{Doc}) = \text{LLM} (p_\text{0},p_\text{A} \oplus \dots \oplus p_\text{D}, \mathcal{Z}_\text{Doc}).
\end{equation}

To construct the latent variable $\mathcal{Z}_\text{Doc}$, dense retrieval models, such as BGE~\cite{bge_embedding}, can be utilized to retrieve query-relevant patents, thereby representing the entire patent corpus $\mathcal{D}$. Specifically, both the query $p_\text{0}$ and document $d_i$ are encoded using Pretrained Language Models (PLMs), such as BERT:
\begin{align}
     h(p_\text{0}) &= \text{BERT} (p_\text{0}),\\
     h(d_i) &= \text{BERT} (d_i).
\end{align}
Then, we aim to maximize the probability $P(\mathcal{Z}_\text{Doc}|p_\text{0}, \mathcal{D})$ associated with the latent variable $\mathcal{Z}_\text{Doc}$ by selecting the top-$k$ patent documents to construct the document set $\mathcal{Z}_\text{Doc} = \{d_1, \dots, d_k\}$. The relevance score between the query patent $p_\text{0}$ and the $i$-th patent document $d_i$ can be computed as follows:
\begin{equation}\label{eq:retrieval_score}
    S(p_\text{0}, d_i) = h(p_\text{0}) \cdot h(d_i).
\end{equation}

\textbf{Memory Graph Augmented Patent Matching.} Then we introduce our memory graph augmented patent matching (\method{}) method, which is built upon the retrieval-augmented framework. It enhances both the patent matching and retrieval processes by respectively integrating the latent variable $\mathcal{Z}_\text{IR}$ and $\mathcal{Z}_\text{Gen}$, which are derived from the memory graph. 

\textit{Patent Document Retrieval.} To enhance the retrieval process, our \method{} incorporates the latent variable $\mathcal{Z}_\text{IR}$ into the retrieval pipeline. Specifically, we calculate the retrieval probability $P(\mathcal{Z}_\text{Doc}|p_\text{0}, \mathcal{D}, \mathcal{Z}_\text{IR})$ by using the expanded query $p_\text{0}^*$:
\begin{equation}\label{eq:retrieval} 
p_\text{0}^* = p_\text{0} \oplus \mathcal{Z}_\text{IR}, 
\end{equation}
where $\mathcal{Z}_\text{IR}$ is a latent variable that can be generated by LLMs:
\begin{equation}
\mathcal{Z}_\text{IR} = \mathcal{T}_\textbf{IR}(V^e (p_\text{0})), 
\end{equation}
where $V^e (p_\text{0}) = \{v^e_1(p_\text{0}), \dots, v^e_n(p_\text{0})\}$ consists of related entities of the query patent $p_\text{0}$. $\mathcal{T}_\textbf{IR}(\cdot)$ is the template to construct the latent variable $\mathcal{Z}_\text{IR}$:
\begin{equation}
\mathcal{T}_\textbf{IR}(V^e (p_\text{0})) = v^e_1(p_\text{0}) \oplus \dots \oplus v^e_n(p_\text{0}), 
\end{equation}
Then $p_\text{0}^*$ is used to compute the relevance score using the same dense retriever as Eq.~\ref{eq:retrieval_score}:
\begin{equation}
S(p_\text{0}^*,d_i) = h(p_\text{0}^*) \cdot h(d_i).
\end{equation}
This relevance score is used to select patent documents that are more closely related to the query patent $p_\text{0}^*$. The selected documents are then utilized to construct $\mathcal{Z}_\text{Doc}$ to provide more precise information for subsequent matching.

\textit{Patent Matching}. Then we leverage the latent variable $\mathcal{Z}_\text{Gen}$ to help LLMs make more accurate predictions:
\begin{equation}\label{eq:generation}
P(\Tilde{y}|p_\text{0}, \mathcal{P}, \mathcal{Z}_\text{Doc}, \mathcal{Z}_\text{Gen}) = \text{LLM} (p_\text{0},p_\text{A} \oplus \dots \oplus p_\text{D}, \mathcal{Z}_\text{Doc}, \mathcal{Z}_\text{Gen}),
\end{equation}
where the latent variable $\mathcal{Z}_\text{Gen}$ can be calculated:
\begin{equation}
\mathcal{Z}_\text{Gen} = \mathcal{T}_\textbf{Gen}(V^o(p_\text{0}), V^o(p_\text{A}), \dots, V^o(p_\text{D})),
\end{equation}
where $V^o(p)=\{ v^o_1(p), ..., v^o_m(p)\}$ contain ontologies that are related with the patent $p$, respectively. The template $\mathcal{T}_\textbf{Gen}(\cdot)$ are defined as follow:
\begin{equation}
\begin{aligned}
 & \mathcal{T}_\textbf{Gen}(V^o(p_\text{0}), V^o(p_\text{A}), \dots, V^o(p_\text{D})) \\&= \text{Query Patent:} V^o(p_\text{0}),\; \text{Option A:} V^o(p_\text{A}),..., \text{Option D:} V^o(p_\text{D}).   
\end{aligned}
\end{equation}

\textit{Summary.} To enhance the performance of retrieval-augmented patent matching process, \method{} introduces the latent variables $\mathcal{Z}_\text{IR}$ and $\mathcal{Z}_\text{Gen}$, aiming to improving both the retrieval (Eq.\ref{eq:retrieval}) and generation (Eq.\ref{eq:generation}) processes. Specifically, \method{} leverages query-related entities $V^e(p)$ as the latent variable $\mathcal{Z}_\text{IR}$ and incorporates ontologies $V^o(p)$, $V^o(p_\text{A}),\dots,V^o(p_\text{D})$ as the latent variable $\mathcal{Z}_\text{Gen}$. In Sec.~\ref{model:architecture}, we provide a detailed explanation of how patent-related entities $V^e(\cdot)$ and ontologies $V^o(\cdot)$ are derived by traversing the memory graph, which is constructed using the parametric memory of LLMs.

\subsection{Traversing the Memory Graph of LLMs}\label{model:architecture}
Given the patent document $p$, we detail the process of obtaining related entities $V^e(p)$ and ontologies $V^o(p)$ by traversing the memory graph. We first introduce the preliminary of patent knowledge graph and then describe our memory graph traversing method. 

\textbf{Preliminary of Patent Knowledge Graph.} For patents, the International Patent Classification (IPC) system serves as a crucial framework for organizing and categorizing patents. It provides a standardized taxonomy for technological innovations, systematically grouping patents by their technical fields~\cite{leydesdorff2008patent}. This classification facilitates efficient comparison and analysis of patents across various regions and industries.

Based on the IPC, we define the patent knowledge graph $G = (V, E)$, comprising nodes $V$ and edges $E$. All nodes are associated by the parametric memory of LLMs, and we leverage LLMs to selectively visit different nodes in an autoregressive decoding way. 
In the patent knowledge graph, there are two kinds of nodes $V$, which include entity nodes ($V^e$) and ontology nodes ($V^o$):
\begin{equation}
    V = V^e \cup V^o.
\end{equation}
Specifically, these entity nodes ($V^e$) represent specific technical concepts, components, or methods found in patents (\textit{e.g.}, ``CMOS sensor''). They are keywords extracted from the given patent. In contrast, the ontology nodes ($V^o$) denote abstract categories, attributes, or relationship types predefined by the IPC system (\textit{e.g.}, ``Image Processing Technology'') for organizing technical concepts in a hierarchical format. This ontology explicitly and systematically defines relationships between these concepts, enabling efficient traversal and enhanced semantic understanding.

\textbf{Graph Traversal Using the Memory of LLMs.} The memory graph originates from the parametric memory of LLMs, which is structured as a network where concepts are interconnected through semantic associations~\cite{collins1975spreading}. All prompts used by \method{} for traversing the memory graph are listed in Table~\ref{tab:prompt}.

To construct the $p$-related entities $V^e(p) = \{v^e_1(p), \dots, v^e_n(p)\}$, we traverse the patent knowledge graph embedded in the parametric memory of LLMs to visit the entity nodes $v^e_k(p)$:
\begin{equation}
P(v^e_k(p)| p, v^e_{1:k-1}(p)) = \text{LLM}(\text{Instruct}_e, p, v^e_{1:k-1}(p)).
\end{equation}
where $\text{Instruct}_e$ is the instruction prompting LLMs to decode the $p$-related entity $v^e_k(p)$ at the $i$-th step. Subsequently, we collect $m$ ontology nodes $V^o(p) = \{ v^o_1(p), ..., v^o_m(p)\}$ based on the given patent $p$ and its related entities $V^e(p)$. To traverse the ontology nodes, we also rely on LLMs to automatically visit the patent-related ones:
\begin{equation}
P(v^o_j(p)| p, v^o_{1:j-1}(p)) = \text{LLM}(\text{Instruct}_o, p, V^e(p), v^o_{1:j-1}(p)),
\end{equation}
where $\text{Instruct}_o$ is the instruction prompting LLMs to traverse the ontology nodes.

\begin{table}[t]
\centering
\small
\caption{Data Statistics of PatentMatch. The International Patent Classification (IPC) information is also provided.}
\label{tab:dataset}
\begin{tabular}{llrr}
\hline
\textbf{IPC} & \textbf{Description} & \textbf{Count} & \textbf{Prop.} \\
\hline
HUM & Human Necessities & 304 & 30.4\% \\
OPER & Performing Operations; & 264 & 26.4\% \\
& Transporting \\
CHEM & Chemistry; Metallurgy & 60 & 6.0\% \\
TEXT & Textiles; Paper & 26 & 2.6\% \\
CONS & Fixed Constructions & 40 & 4.0\% \\
MECH & Mechanical Engineering; Lighting; & 100 & 10.0\% \\
& Heating, Weapons \\
PHYS & Physics & 160 & 16.0\% \\
ELEC & Electricity & 46 & 4.6\% \\
\hline
\textbf{Total} & & 1,000 & 100\% \\
\hline
\end{tabular}%
\end{table}
\section{Experimental Methodology}
In this section, we describe the dataset, evaluation metrics, baselines, and implementation details used in our experiments.

\textbf{Dataset.} We evaluate the patent matching capabilities of different methods using the PatentMatch dataset~\cite{ni2024mozip}, which consists of 1,000 patent matching questions derived from authentic patent documents. As shown in Table~\ref{tab:dataset}, the dataset is evenly distributed across two languages (Chinese and English) and covers all eight sections defined by the International Patent Classification (IPC). To conduct the retrieval-augmented generation (RAG) methods, we construct a patent corpus comprising 300,000 patent documents, using the BGE-Base model~\cite{bge_embedding} as the retriever. The Top-3 ranked retrieved patents are utilized as input for RAG-based methods.

\begin{table}[!t]
\centering
\setlength{\tabcolsep}{5pt}
\caption{Overall Performance. We show the performance of \method{} and baseline methods on the PatentMatch datasets. $\dagger$ and $\ddagger$ indicate statistically significant improvements over Vanilla LLM and RAG Models, respectively.}
\label{tab:overall}
\begin{tabular}{lccc}
\hline
\textbf{Method} &  \textbf{English} & \textbf{Chinese} &  \textbf{Avg.} \\
\hline
\rowcolor{gray!8}\multicolumn{4}{l}{\textbf{Vanilla LLM}} \\
MoZi$_\textsc{7B}$~\cite{ni2024mozip} & 25.8 & 29.0 & 27.4 \\
PatentGPT$_\textsc{1.5B}$~\cite{ren2024patentgpt}&  26.2 & - & -  \\
PatentGPT-1.0-Dense$_\textsc{70B}$~\cite{bai2024patentgpt} &  66.2 & 72.0 & 69.1 \\
\cdashline{1-4}
Llama-3.1-Instruct$_\textsc{8B}$ & 43.0 & 51.0 & 47.0 \\
Qwen2-Instruct$_\textsc{7B}$ & 31.4 & 47.0 & 39.2 \\
GLM-4-Chat$_\textsc{9B}$ & 66.4 & 65.0 & 65.7 \\
Qwen2.5-Instruct$_\textsc{14B}$ & 63.8 & 70.0 & 66.9 \\
\hline
\rowcolor{gray!8}\multicolumn{4}{l}{\textbf{Chain-of-Thought (CoT)}}  \\
Llama-3.1-Instruct$_\textsc{8B}$ & 44.4 & 54.0 & 49.2 \\
Qwen2-Instruct$_\textsc{7B}$ & 32.8 & 49.2 & 41.0 \\
GLM-4-Chat$_\textsc{9B}$ & 68.0 & 65.6 & 66.8 \\
Qwen2.5-Instruct$_\textsc{14B}$ & 64.0 & 71.2 & 67.6 \\
\hline
\rowcolor{gray!8}\multicolumn{4}{l}{\textbf{Retrieval-Augmented Generation (RAG)}} \\ 
Llama-3.1-Instruct$_\textsc{8B}$ & 49.4$^\dagger$ & 45.2 & 47.3 \\
Qwen2-Instruct$_\textsc{7B}$ & 49.2$^\dagger$ & 68.2$^\dagger$ & 58.7 \\
GLM-4-Chat$_\textsc{9B}$ & 75.8$^\dagger$ & 69.4$^\dagger$ & 72.6 \\
Qwen2.5-Instruct$_\textsc{14B}$ & 70.8$^\dagger$ & 64.2 & 67.5 \\
\hline
\rowcolor{gray!8}\multicolumn{4}{l}{\textbf{\method{}}} \\ 
Llama-3.1-Instruct$_\textsc{8B}$ & 66.0$^{\dagger\ddagger}$ & 64.6$^{\dagger\ddagger}$ & 65.3 \\
Qwen2-Instruct$_\textsc{7B}$ & 62.6$^{\dagger\ddagger}$ & 71.4$^{\dagger\ddagger}$ & 67.0 \\
GLM-4-Chat$_\textsc{9B}$ & \textbf{82.8}$^{\dagger\ddagger}$ & \textbf{80.8}$^{\dagger\ddagger}$ & \textbf{81.8} \\
Qwen2.5-Instruct$_\textsc{14B}$ & 75.8$^{\dagger\ddagger}$ & 75.0$^{\dagger\ddagger}$ & 75.4 \\
\hline
\end{tabular}
\end{table}
\textbf{Evaluation Metrics.} 
Following prior work~\cite{ni2024mozip,bai2024patentgpt,ren2024patentgpt}, we adopt Accuracy as the primary metric to assess the effectiveness of various methods for patent matching. Statistic significances are tested by permutation test with P$<0.05$.

\textbf{Baselines.} We compare \method{} against several baseline models, including patent-specific fine-tuned LLMs, chain-of-thought reasoning~\cite{wei2022chain}, and various retrieval-augmented generation (RAG) methods~\cite{lewis2020retrieval}. We employ three domain-specific models as baselines, including MoZi\cite{ni2024mozip}, PatentGPT\cite{ren2024patentgpt}, and PatentGPT-1.0-Dense~\cite{bai2024patentgpt}, all of which are trained on patent corpora. Chain-of-thought reasoning~\cite{wei2022chain} aims to prompt LLMs to analyze the patent documents step-by-step, enabling a deeper understanding when generating answers. RAG-based methods leverage retrieved patents as contextual information, prompting LLMs to identify the patent with the most similar innovations.

\textbf{Implementation Details.} In our experiments, we employ different LLMs as the foundation model of \method{}, including Llama-3.1-Instruct$_\textsc{8B}$~\cite{dubey2024llama}, Qwen2-Instruct$_\textsc{7B}$~\cite{yang2024qwen2}, GLM-4-Chat$_\textsc{9B}$~\cite{glm2024chatglm}, and Qwen2.5-Instruct$_\textsc{14B}$~\cite{qwen2.5}. Specifically, we use the Transformers toolkit~\cite{wolf2020transformers} to implement Llama, GLM, and Qwen2 models. For Qwen2.5 model, we utilize the Dashscope\footnote{\url{https://dashscope.aliyun.com/}} SDK to access the Qwen API. During retrieval, we implement BGE-Base~\cite{bge_embedding} using the FlagEmbedding toolkit\footnote{\url{https://github.com/FlagOpen/FlagEmbedding}}.
\begin{table*}[!t]
\centering
\caption{Ablation Study. We evaluate the effectiveness of different modules in our \method{} method on the PatentMatch dataset across IPC sections. The Llama-3.1-Instruct$_\textsc{8B}$, Qwen2-Instruct$_\textsc{7B}$, GLM-4-Chat$_\textsc{9B}$, and Qwen2.5-Instruct$_\textsc{14B}$ models are used for evaluation. More details of IPC sections are shown in Table~\ref{tab:overall}.}
\label{tab:ablation}
\begin{tabular}{lccccccccc}
\hline
\multirow{2}{*}{\textbf{Method}} & \multicolumn{8}{c}{\textbf{IPC Sections}} & \multirow{2}{*}{\textbf{Avg.}} \\

 & HUM & OPER & CHEM & TEXT & CONS & MECH & PHYS & ELEC & \\
\hline
\rowcolor{gray!8}\multicolumn{10}{l}{\textbf{Llama-3.1-Instruct$_\textsc{8B}$}} \\
Vanilla LLM & 42.1 & 37.9 & 53.3 & 50.0 & 45.0 & 53.0 & 46.9 & 43.5 & 46.5\\
Vanilla RAG & 50.0 & 50.0 & 45.0 & 65.4 & 40.0 & 54.0 & 47.5 & 37.0 & 48.6\\
\method{} (Only $\mathcal{Z}_\text{IR}$) & 48.4 & 49.2 & 56.7 & \textbf{84.6} & 40.0 & 57.0 & 54.4 & 47.8 & 54.8\\
\method{} (Only $\mathcal{Z}_\text{Gen}$) & 60.9 & 59.5 & 61.9 & 69.2 & 65.0 & \textbf{73.0} & 61.3 & \textbf{63.0} & 64.2\\
\method{} (All) & \textbf{63.2} & \textbf{62.5} & \textbf{75.0} & 73.1 & \textbf{70.0} & \textbf{73.0} & \textbf{63.8} & \textbf{63.0} & \textbf{67.9}\\
\hline
\rowcolor{gray!8}\multicolumn{10}{l}{\textbf{Qwen2-Instruct$_\textsc{7B}$}} \\
Vanilla LLM & 38.5 & 36.7 & 41.7 & 26.9 & 42.5 & 46.0 & 54.4 & 47.8 & 41.8\\
Vanilla RAG & 58.2 & 59.5 & 65.0 & \textbf{61.5} & 57.5 & 56.0 & 58.8 & \textbf{54.3} & 58.9\\
\method{} (Only $\mathcal{Z}_\text{IR}$) & 60.9 & 58.3 & 66.7 & \textbf{61.5} & 60.0 & 54.0 & 55.6 & \textbf{54.3} & 58.9\\
\method{} (Only $\mathcal{Z}_\text{Gen}$) & \textbf{68.1} & 68.6 & 75.0 & 50.0 & \textbf{70.0} & 61.0 & 58.1 & 50.0 & 62.6\\
\method{} (All) & \textbf{68.1} & \textbf{68.9} & \textbf{76.7} & \textbf{61.5} & \textbf{70.0} & \textbf{68.0} & \textbf{61.9} & 52.5 & \textbf{65.9}\\
\hline
\rowcolor{gray!8}\multicolumn{10}{l}{\textbf{GLM-4-Chat$_\textsc{9B}$}} \\
Vanilla LLM & 66.4 & 67.0 & 73.3 & 61.5 & 77.5 & 52.0 & 63.8 & 54.3 & 64.5\\
Vanilla RAG & 75.0 & 73.5 & 68.3 & 76.9 & 77.5 & 71.0 & 69.4 & 65.2 & 71.8\\
\method{} (Only $\mathcal{Z}_\text{IR}$) & 77.0 & 72.7 & 73.3 & 76.9 & 77.5 & 71.0 & 70.6 & 67.4 & 73.3\\
\method{} (Only $\mathcal{Z}_\text{Gen}$) & 82.2 & \textbf{79.9} & 76.7 & 80.8 & 92.5 & 85.0 & 77.5 & 69.6 & 80.5\\
\method{} (All) & \textbf{82.9} & \textbf{79.9} & \textbf{80.0} & \textbf{92.3} & \textbf{95.0} & \textbf{86.0} & \textbf{78.1} & \textbf{73.9} & \textbf{83.5}\\
\hline
\rowcolor{gray!8}\multicolumn{10}{l}{\textbf{Qwen2.5-Instruct$_\textsc{14B}$}} \\
Vanilla LLM & 58.6 & 69.3 & 76.7 & \textbf{88.5} & 70.0 & 67.0 & 71.9 & 63.0 & 70.6\\
Vanilla RAG & 62.5 & 74.2 & 75.0 & 61.5 & 57.5 & 64.0 & 71.3 & 58.7 & 65.6\\
\method{} (Only $\mathcal{Z}_\text{IR}$) & 64.1 & 73.5 & 75.0 & 76.9 & 65.0 & 70.0 & 71.3 & 58.7 & 69.3\\
\method{} (Only $\mathcal{Z}_\text{Gen}$) & 70.4 & 78.4 & \textbf{85.0} & \textbf{88.5} & 67.5 & 70.0 & \textbf{72.5} & 63.0 & 74.4\\
\method{} (All) & \textbf{72.4} & \textbf{78.8} & \textbf{85.0} & \textbf{88.5} & \textbf{77.5} & \textbf{73.0} & \textbf{72.5} & \textbf{69.6} & \textbf{77.1}\\
\hline
\end{tabular}
\end{table*}
\section{Evaluation Result}
In this section, we begin by evaluating the overall performance of \method{} and conducting ablation studies. Next, we investigate the effectiveness of \method{} in utilizing retrieved patents and its impact on the process of making patent-matching decisions. Finally, some case studies are shown.

\subsection{Overall Performance\label{eval:overall}}
The patent matching performance of different models is presented in Table~\ref{tab:overall}. Several LLMs are compared in our experiments, including general LLMs and patent-specific LLMs.

Among various LLMs, our experiments demonstrate that general LLMs also exhibit competitive patent-matching performance compared to the LLMs that are finetuned on well-curated patent datasets. The Chain-of-Thought (CoT) approach enables LLMs to better analyze the similarity between query patents and candidate patents, achieving approximately a 1\% improvement over Vanilla LLM. Unlike CoT methods, RAG models show much better patent-matching performance by providing query-related patents. These retrieved patent documents contain matched keywords, which provide some references to help LLMs differentiate between confusing patents. However, RAG models show limited effectiveness when using both Llama-3.1-Instruct$_\textsc{8B}$ and Qwen2.5-Instruct$_\textsc{14B}$ as foundation models. The primary reason might be that these retrieved patent documents include some noise, which misleads the LLMs during the patent-matching process.

Building upon the RAG framework, \method{} achieves significant improvements in patent matching performance, with over a 10\% enhancement, highlighting its effectiveness. 
\method{} equips LLMs with both entities and ontologies by traversing the memory graph of LLMs, which serves as hints to guide LLMs in capturing matching signals among query patents, candidate patents and retrieved patent documents. 
Notably, \method{} also shows significant improvements over baseline models when employing Llama-3.1-Instruct$_\textsc{8B}$ and Qwen2.5-Instruct$_\textsc{14B}$ to construct RAG models, emphasizing the generalization capability of our \method{} approach. 
Furthermore, with the integration of \method{}, GLM-4-Chat$_\textsc{9B}$ achieves the best performance among all models, revealing the potential for smaller-scale LLMs to deliver superior patent matching performance compared to larger counterparts.

\subsection{Ablation Study\label{eval:ablation}}
As shown in Table~\ref{tab:ablation}, we conduct ablation studies to analyze the contributions of the latent variables, $\mathcal{Z}_\text{IR}$ and $\mathcal{Z}_\text{Gen}$, within the RAG framework. The PatentMatch dataset is divided into eight categories based on the International Patent Classification (IPC) to evaluate the patent matching effectiveness of models across different domains.

In this experiment, we evaluate three ablation models: \method{} (Only $\mathcal{Z}_\text{IR}$), \method{} (Only $\mathcal{Z}_\text{Gen}$), and \method{} (All). Specifically, the \method{} (Only $\mathcal{Z}_\text{IR}$) model leverages only the latent variable $\mathcal{Z}_\text{IR}$ to enhance the retrieval process within the RAG-based patent matching framework. Conversely, the \method{} (Only $\mathcal{Z}_\text{Gen}$) model utilizes only the latent variable $\mathcal{Z}_\text{Gen}$ to enhance the generator module for producing patent matching results. \method{} (All) integrates both latent variables, $\mathcal{Z}_\text{IR}$ and $\mathcal{Z}_\text{Gen}$, to improve the entire RAG framework. The experiments employ Llama-3.1-Instruct$_\textsc{8B}$, Qwen2-Instruct$_\textsc{7B}$, GLM-4-Chat$_\textsc{9B}$, and Qwen2.5-Instruct$_\textsc{14B}$ as the foundation LLMs.

Compared to the Vanilla RAG method, \method{} (Only $\mathcal{Z}_\text{IR}$) achieves an average improvement of 3.8\% in matching performance. Notably, \method{} (Only $\mathcal{Z}_\text{IR}$) also mitigates the performance degradation observed in Vanilla RAG when implemented using Qwen2.5-Instruct$_\textsc{14B}$ as the backbone model. These results show the effectiveness of the latent variable $\mathcal{Z}_\text{IR}$ in enhancing the performance of RAG-based patent matching models. The latent variable $\mathcal{Z}_\text{IR}$ incorporates several entities that enable the retriever to return more accurate evidence for patent matching. In comparison, \method{} (Only $\mathcal{Z}_\text{Gen}$) achieves even greater improvements, with a 9.2\% increase in patent matching performance over Vanilla RAG models, demonstrating that the primary contribution of \method{} stems from the latent variable $\mathcal{Z}_\text{Gen}$. Furthermore, \method{} (All) achieves the best performance by integrating both latent variables, $\mathcal{Z}_\text{IR}$ and $\mathcal{Z}_\text{Gen}$, revealing the effectiveness of our \method{}.

Further experiments evaluate the patent matching effectiveness of different models across different IPC types. Compared with \method{} (Only $\mathcal{Z}_\text{Gen}$), \method{} (All) achieves greater improvements in types such as CHEM, CONS, and MECH. These categories require the retriever strengths of these domain-specific terminologies for matching the query patent and its related patents. Additionally, compared with \method{} (Only $\mathcal{Z}_\text{IR}$), \method{} (All) achieves more consistent improvements, showing the effectiveness of $\mathcal{Z}_\text{Gen}$ in guiding LLMs to extract key point information from retrieved evidence and capture matching signals between the query and candidate patents. As the backbone model scales up, \method{} (All) maintains robust performance across different IPC types, further demonstrating the robustness of our \method{}.

\subsection{The Effectiveness of \method{} in Assisting LLMs with Utilizing Retrieved Patents\label{eval:external}}
As shown in Table~\ref{tab:rag-comparison}, we first use the raw query patent to search related patents and then evaluate how the latent variable $\mathcal{Z}_\text{Gen}$ helps LLMs use these retrieved patents.

In this experiment, the PatentMatch dataset is divided into three testing scenarios with the following proportions of the total dataset: Hit-Choice (8.8\%), Miss-Choice (91.2\%), and Mem-Choice (Llama-3.1-Instruct$_\textsc{8B}$: 47\%, Qwen2-Instruct$_\textsc{7B}$: 39.2\%, GLM-4-Chat$_\textsc{9B}$: 65.7\%, and Qwen2.5-Instruct$_\textsc{14B}$: 66.9\%). In the Hit-Choice scenario, all testing instances include the golden patent in the retrieved patents. On the contrary, the Miss-Choice scenario represents cases where the golden patent is absent from the retrieved evidence set. The Mem-Choice scenario consists of instances that can be accurately answered by Vanilla LLM, whereas RAG-based methods may produce incorrect answers. This evaluation scenario focuses more on investigating how $\mathcal{Z}_\text{Gen}$ helps resolve noise derived from these retrieved patents.

As shown in the evaluation results, the Vanilla RAG models exhibit improvements in the Hit-Choice and Miss-Choice testing scenarios, as the retriever provides query-related patents enhancing the probability of selecting the correct answer. It is evident that more accurate retrieval evidence can improve the RAG performance, making it achieve much better performance in the Hit-Choice scenario. Compared with Vanilla RAG, \method{} (Only $\mathcal{Z}_\text{Gen}$) achieves more improvements in the Miss-Choice scenario. This indicates that the latent variable $\mathcal{Z}_\text{Gen}$ helps LLMs better capture critical information from the retrieved patents and utilize it effectively during the patent matching process. In the Mem-Choice scenario, these RAG models are misled by the retrieved patents, resulting in a performance decrease. \method{} (Only $\mathcal{Z}_\text{Gen}$) mitigates this issue by narrowing the performance gap by approximately 4\%. This demonstrates that the latent variable $\mathcal{Z}_\text{Gen}$, derived from the memory graph, aids LLMs in resolving conflicts between external knowledge and internal memory, thereby improving patent matching outcomes.

\begin{table}[t]
\small
\centering
\setlength{\tabcolsep}{4pt} 
\caption{Patent Matching Performance of the RAG Models Incorporating the Latent Variable $\mathcal{Z}_\text{Gen}$. We evaluate the performance of Vanilla LLM, Vanilla RAG, and \method{} (Only $\mathcal{Z}_\text{Gen}$) to assess their ability to leverage external knowledge.}
\label{tab:rag-comparison}
\begin{tabular}{lccc}
\hline
\textbf{Method} & \textbf{Hit-Choice} & \textbf{Miss-Choice} & \textbf{Mem-Choice} \\
\hline
\rowcolor{gray!8}\multicolumn{4}{l}{\textbf{Llama-3.1-Instruct$_\textsc{8B}$}} \\
Vanilla LLM & 51.1 & 41.9 & 100.0 \\
\cdashline{1-4}
Vanilla RAG & 84.1 & 48.7 & 70.0 \\
\method{} (Only $\mathcal{Z}_\text{Gen}$) & \textbf{86.4} & \textbf{61.5} & \textbf{74.9} \\
\hline
\rowcolor{gray!8}\multicolumn{4}{l}{\textbf{Qwen2-Instruct$_\textsc{7B}$}} \\
Vanilla LLM & 43.2 & 38.8 & 100.0 \\
\cdashline{1-4}
Vanilla RAG & 79.5 & 56.7 & 77.8 \\
\method{} (Only $\mathcal{Z}_\text{Gen}$) & \textbf{89.8} & \textbf{64.8} & \textbf{79.3} \\
\hline
\rowcolor{gray!8}\multicolumn{4}{l}{\textbf{GLM-4-Chat$_\textsc{9B}$}} \\
Vanilla LLM & 84.1 & 63.9 & 100.0 \\
\cdashline{1-4}
Vanilla RAG & 85.2 & 71.4 & 88.9 \\
\method{} (Only $\mathcal{Z}_\text{Gen}$) & \textbf{89.8} & \textbf{81.0} & \textbf{92.2} \\
\hline
\rowcolor{gray!8}\multicolumn{4}{l}{\textbf{Qwen2.5-Instruct$_\textsc{14B}$}} \\
Vanilla LLM & 77.3 & 65.9 & 100.0 \\
\cdashline{1-4}
Vanilla RAG & 88.6 & 65.5 & 84.5 \\
\method{} (Only $\mathcal{Z}_\text{Gen}$) & \textbf{90.9} & \textbf{74.0} & \textbf{89.2} \\
\hline
\end{tabular}
\end{table}

\subsection{Evaluating the Impact of \method{} on the Reasoning Process of Patent Matching\label{eval:impact}}
\begin{figure}[!t]
    \centering
    \subfigure[Perplexity Scores of Predicting the Ground Truth Answers.]{
        \label{fig:two_scenarios}
        \includegraphics[width=0.45\linewidth]{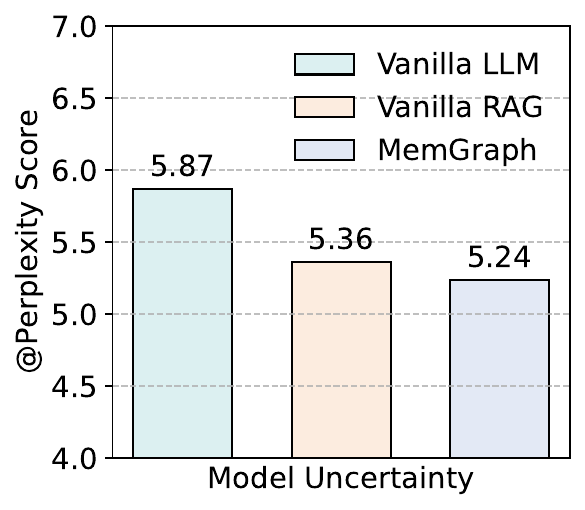}
    }
    \subfigure[The Win Rate of Various Models in Generating Patent Matching Reasoning Outcomes.]{
        \label{fig:reasoning}
        \includegraphics[width=0.45\linewidth]{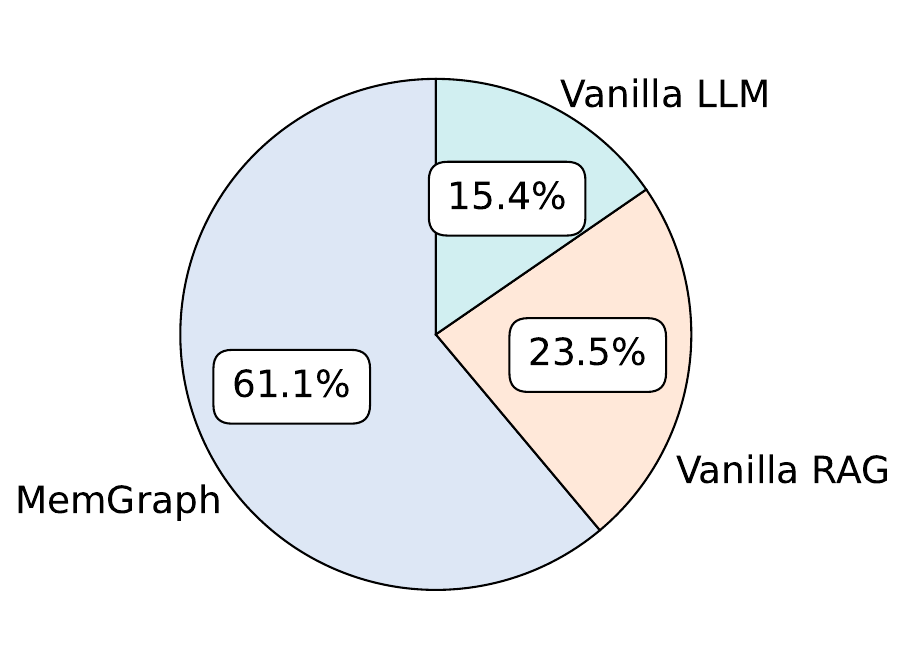  }
    }
    \caption{Evaluation Results of Different Models in Patent Matching Predictions. Figure~\ref{fig:two_scenarios} illustrates the model uncertainty when generating the patent matching results. Figure~\ref{fig:reasoning} demonstrates the quality of the reasoning process during patent matching. All models are implemented by using Llama-3.1-Instruct$_\textsc{8B}$ as the backbone model.}
    \label{fig:decision}
\end{figure}

In this subsection, we evaluate the effectiveness of \method{} in improving the reasoning process of LLMs for the patent matching task. As illustrated in Figure~\ref{fig:decision}, we assess the performance of Vanilla LLM, Vanilla RAG models, and \method{} by considering both model uncertainty and reasoning accuracy while producing the choice prediction for the patent matching task.

\textbf{Model Uncertainty.} As shown in Figure~\ref{fig:two_scenarios}, we present the perplexity scores of different models when generating the ground truth options. The perplexity score reflects the model's uncertainty in making predictions~\cite{bengio2000neural}. A lower score indicates that the model is more confident in selecting the correct option.

As shown in the evaluation results, the RAG model significantly reduces the perplexity score of Vanilla LLM when producing the ground truth option. This demonstrates that retrieved patents help guide LLMs, boosting their confidence in selecting the correct options. Moreover, \method{} brings a further decrease to the perplexity of Vanilla RAG models when predicting the ground truth option, showing the effectiveness of \method{} in steering LLMs toward more accurate predictions. Such a phenomenon illustrates that a more thorough exploration of memory knowledge of LLMs is necessary for patent matching, making LLMs fully utilize the retrieved patents to improve decision-making confidence. By incorporating information from the memory graph, \method{} enables LLMs to capture more critical semantics from relevant patents and use them to produce more reliable prediction outcomes.

\textbf{Reasoning Quality.} As shown in Figure~\ref{fig:reasoning}, we further evaluate the quality of the reasoning process in patent matching using the entire PatentMatch dataset, which contains 1,000 data instances.

During evaluation, we utilize the Llama-3.1-Instruct$_\textsc{8B}$ model as the backbone and evaluate Vanilla LLM, Vanilla RAG, and \method{} models on both English and Chinese patent matching datasets. To assess the reasoning quality of different models, we prompt each model to generate chain-of-thought outcomes for the patent matching task. Subsequently, GPT-4o~\cite{achiam2023gpt,hurst2024gpt} is employed as the judge to evaluate the reasoning process quality.

As shown in the evaluation results, both the Vanilla LLM model and the Vanilla RAG model achieve win rates of 15.4\% and 23.5\%, respectively, across all testing cases. This highlights that retrieved evidence can assist LLMs in producing more convincing reasoning outcomes. Our \method{} method demonstrates its effectiveness by securing the preference of GPT-4o in 61.1\% of testing cases. This substantial improvement further underscores that the internal reasoning processes of LLMs can significantly enhance the reasoning capabilities of RAG models, which is also observed in prior RAG models~\cite{xu2024activerag}. Specifically, \method{} traverses the parametric memory to retrieve related ontologies for both the query patent and the candidate patent. This approach mimics expert human cognition, enabling LLMs to simultaneously account for detailed technical nuances and broader technological categories. By incorporating these ontologies, LLMs gain a deeper understanding of both the query patent and the candidate patent by using the associations among ontologies, which improves the depth and interpretability of the reasoning process during the patent-matching process.

\begin{table*}
\centering
\small
\captionsetup{font=small,labelfont=bf}
\caption{Case Study. We present one case to analyze the behaviors of different models in the patent matching task. We highlight these potential matched phrases between the query patent and the candidate/retrieved patent for evaluation. Different colors are used to annotate these matched phrases for each model: \sethlcolor{lightyellow}\hl{Green} for Vanilla LLM, \sethlcolor{lightgreen}\hl{Orange} for Vanilla RAG, and \sethlcolor{lightblue}\hl{Blue} for \method{}.}
\label{tab:case_study}
\begin{tabular}{>{\raggedright\arraybackslash}p{0.98\textwidth}}
\toprule
\multicolumn{1}{c}{\textbf{Input Data (No Options)}}  \\
\textbf{Instruction:} Please select the most similar patent number from A, B, C and D. Which number is? \\
\textbf{Query Patent:} A \sethlcolor{lightblue}\hl{sauce-like fluid fish feed} and a \sethlcolor{lightyellow}\hl{preparation method} thereof. The \sethlcolor{lightyellow}\hl{raw materials} of the sauce-like fluid fish feed include \sethlcolor{lightblue}\hl{imported fish meal}, \sethlcolor{lightblue}\hl{Antarctic krill meal}, \sethlcolor{lightblue}\hl{squid meal}, \sethlcolor{lightblue}\hl{$\alpha$-starch}, \sethlcolor{lightyellow}\hl{vitamins}, minerals, \sethlcolor{lightblue}\hl{fish oil}, \sethlcolor{lightblue}\hl{soybean lecithin oil}, \sethlcolor{lightblue}\hl{betaine}, \sethlcolor{lightblue}\hl{edible glue}, \sethlcolor{lightblue}\hl{water retaining agent}, preservative, chitosan, and \sethlcolor{lightgreen}\hl{digestive enzyme}. The preparation method includes \sethlcolor{lightgreen}\hl{mixing and crushing} the raw materials, performing \sethlcolor{lightgreen}\hl{enzymolysis}, heating... \\
\midrule
\multicolumn{1}{c}{\textbf{Vanilla LLM Choice Result}}  \\
\textbf{Chosen:} \textcolor{lightred}{(Incorrect)} \\
\textbf{D.} A psoralen polymer nanoparticle \sethlcolor{lightyellow}\hl{preparation method} thereof... The method uses \sethlcolor{lightyellow}\hl{raw materials} including psoralen, polylactide-glycolide, \sethlcolor{lightyellow}\hl{soybean lecithin} and polyethylene glycol 1000 \sethlcolor{lightyellow}\hl{vitamin} E succinate... The encapsulation efficiency of the psoralen polymer nanoparticle preparation can be further improved by optimizing the preparation conditions... \\
\midrule
\multicolumn{1}{c}{\textbf{Vanilla RAG Choice Result}} \\ 
\textbf{Retrieved Passages:} \\
(1) A method for preparing a \sethlcolor{lightgreen}\hl{fish bait preservative}... uses a combined strain of yeast, lactic acid bacteria and \sethlcolor{lightgreen}\hl{enzymes}... performs \sethlcolor{lightgreen}\hl{fermentation} and cultivation... \\
(2) A method for making \sethlcolor{lightgreen}\hl{fish products}, comprising... \sethlcolor{lightgreen}\hl{crushing} into meat paste... \sethlcolor{lightgreen}\hl{steaming and sterilizing}... \\
(3) A \sethlcolor{lightgreen}\hl{capsule feed} comprises... \sethlcolor{lightgreen}\hl{fish meal}, \sethlcolor{lightgreen}\hl{fish oil}, \sethlcolor{lightgreen}\hl{phospholipid oil}... \\
\textbf{Chosen:} \textcolor{lightred}{(Incorrect)} \\
\textbf{B.} A non-meat source yeast extract with a strong meat flavor... The preparation method comprises \sethlcolor{lightgreen}\hl{mixing} yeast extract with \sethlcolor{lightgreen}\hl{enzymes}, performing \sethlcolor{lightgreen}\hl{fermentation} at controlled temperature and pH conditions, followed by \sethlcolor{lightgreen}\hl{sterilization} and drying to obtain a powder with strong meat flavor characteristics... \\
\midrule
\multicolumn{1}{c}{\textbf{MemGraph Choice Result}} \\ 
\textbf{Retrieved Passages:} \\
(1) An \sethlcolor{lightblue}\hl{aquaculture feed} in micro-particle form... includes \sethlcolor{lightblue}\hl{fish meal}, \sethlcolor{lightblue}\hl{krill powder}, and \sethlcolor{lightblue}\hl{squid extract}... \\
(2) A \sethlcolor{lightblue}\hl{feed to improve fish immunity}... comprises \sethlcolor{lightblue}\hl{fish meal}, \sethlcolor{lightblue}\hl{Antarctic krill}, and \sethlcolor{lightblue}\hl{marine phospholipids}... \\
(3) A \sethlcolor{lightblue}\hl{fluid fish feed} with enhanced digestibility... contains \sethlcolor{lightblue}\hl{imported fish meal} and \sethlcolor{lightblue}\hl{squid meal}... \\
\textbf{Entity:} \\
\sethlcolor{lightblue}\hl{Sauce-like fluid fish feed}; \sethlcolor{lightblue}\hl{Imported fish meal}; \sethlcolor{lightblue}\hl{Antarctic krill meal}; \sethlcolor{lightblue}\hl{squid meal}; 
\sethlcolor{lightblue}\hl{$\alpha$-starch}; 
\sethlcolor{lightblue}\hl{Fish oil}; \sethlcolor{lightblue}\hl{Soybean lecithin oil};
\sethlcolor{lightblue}\hl{Betaine}; 
\sethlcolor{lightblue}\hl{Edible gum}; 
\sethlcolor{lightblue}\hl{Water retaining agent} \\
\textbf{Ontology:} \\
\underline{Query Patent:} \sethlcolor{lightblue}\hl{Food processing > Mixtures > Fluid fish feed}; \underline{Option A:} \sethlcolor{lightblue}\hl{Food processing > Mixtures > Fish feed}; \underline{Option B:} Food processing > Yeast extract > Meat-flavored yeast; \underline{Option C:} Food processing > Mixtures > Meal replacement powders; \underline{Option D:} Pharmaceutical preparations > Nanoparticle preparations > Psoralen polymers \\
\textbf{Chosen:} \textbf{\textcolor{red}{(Correct)}} \\
\textbf{A.} An \sethlcolor{lightblue}\hl{aquaculture feed} composition comprising \sethlcolor{lightblue}\hl{fish meal}, \sethlcolor{lightblue}\hl{krill meal}, and \sethlcolor{lightblue}\hl{squid meal} as primary protein sources. The feed is processed into a \sethlcolor{lightblue}\hl{fluid form} to enhance digestibility and nutrient absorption... The composition maintains optimal \sethlcolor{lightblue}\hl{protein-to-lipid ratios} suitable for marine species... \\
\bottomrule
\end{tabular}
\end{table*}
\subsection{Case Study\label{eval:case}}
Finally, we randomly sample one case in Table~\ref{tab:case_study} to show the patent matching effectiveness of different models. The query patent describes the ``sauce-like fluid fish feed'', and we analyze how each model conducts the prediction during matching patents.

As shown in the selected case, both Vanilla LLM and Vanilla RAG exhibit limited matching patterns in their performance analysis. The selected patent of the Vanilla LLM model predominantly matches some general terms in the patent such as ``preparation method'', ``raw materials'', and ``vitamins''. This leads to an incorrect match with Option D (Psoralen polymer nanoparticle patent), which is unrelated to the query patent. The primary reason for this mismatch maybe the procedural similarities in preparation methods between the two patents and their shared use of soy lecithin and vitamins as ingredients. Compared with the Vanilla LLM model, Vanilla RAG shows more reliable performance by incorporating additional query-related keywords from retrieved patents during the patent-matching process. These retrieved passages (highlighted in green) emphasize critical information, including ``enzymes'', ``fermentation'', ``fish products'', and ``sterilization'', which are essential clues for accurately matching the query patent with the ground truth patent. However, the retrieved passages overemphasized processing techniques such as ``enzymatic treatment'' and ``fermentation'', leading to an erroneous match with Option B (a non-meat-derived yeast extract patent with similar processing procedures).

Different from both Vanilla LLM and Vanilla RAG models, our \method{} demonstrates its effectiveness by integrating patent-related entities and ontologies into the RAG framework. For the retrieval results, these extracted entities effectively expand the raw query, facilitating the retrieval of patents that are more closely related to the query patent and reducing noise from the retrieval stage. When generating the patent matching results, the ontologies provided by \method{} accurately identify the classification paths between the query patent and its correct matches (\textit{e.g.}, ``Food Processing > Mixtures > Fish Feed''). These ontologies further enhance the understanding of patent semantics and calibrate LLMs to focus more on these ontology related keywords during the patent-matching process, thereby refining the matching clues between the query patent and the ground-truth patent.

\section{Conclusion}

In this paper, we propose \method{}, a novel framework to enhance the patent matching capability of LLMs by traversing the memory graph derived from their parametric memory.  Our experiments demonstrate that \method{} enhances both the retriever and generator by incorporating entities and ontologies, and consistently outperforms existing methods across various technical fields. Further analysis reveals that \method{} reduces prediction uncertainty and improves reasoning quality, offering a promising approach for more effective and interpretable patent matching.

\begin{acks}
This work is partly supported by the Natural Science Foundation of China under Grant (No. 62206042 and No. 62461146205), the Joint Funds of Natural Science Foundation of Liaoning Province (No. 2023-MSBA-081), and the Fundamental Research Funds for the Central Universities under Grant (No. N2416012).
\end{acks}

\bibliographystyle{ACM-Reference-Format}
\balance
\bibliography{custom}


\end{document}